\documentstyle[12pt]{article}

 at 10truept

\def\ga{\gamma}
\def\de{\delta}
\def\ep{\epsilon}

\def\la{\lambda}

\def\ph{\phi}

\def\ch{\chi}
\def\ps{\psi}

\def\Ga{\Gamma}
\def\De{\Delta}

\def\La{\Lambda}

\def\cl{{\cal L}}
\def\fr#1#2{{{#1} \over {#2}}}
\def\prt{\partial}

\def\vev#1{\langle {#1}\rangle}

\def\frac#1#2{{\textstyle{{#1}\over {#2}}}}

\def\lsim{\mathrel{\rlap{\lower4pt\hbox{\hskip1pt$\sim$}}
    \raise1pt\hbox{$<$}}}
\def\gsim{\mathrel{\rlap{\lower4pt\hbox{\hskip1pt$\sim$}}
    \raise1pt\hbox{$>$}}}
\def\sqr#1#2{{\vcenter{\vbox{\hrule height.#2pt
         \hbox{\vrule width.#2pt height#1pt \kern#1pt
         \vrule width.#2pt}
         \hrule height.#2pt}}}}

\def\Re{\hbox{Re}\,}
\def\Im{\hbox{Im}\,}

\newcommand{\beq}{\begin{equation}}
\newcommand{\eeq}{\end{equation}}
\newcommand{\bea}{\begin{eqnarray}}
\newcommand{\eea}{\end{eqnarray}}
\newcommand{\rf}[1]{(\ref{#1})}

\renewenvironment{thebibliography}[1]
 { \rm
   \begin{list}{\arabic{enumi}.}
    {\usecounter{enumi} \setlength{\parsep}{0pt}
     \setlength{\itemsep}{3pt} \settowidth{\labelwidth}{#1.}
     \sloppy
    }}{\end{list}}

\begin{document}

\baselineskip=16pt
\begin{flushright}
{IUHET 357\\}
{March 1997\\}
\end{flushright}
\vglue 1.0 truein 

\begin{flushleft}
{\bf CPT VIOLATION, STRINGS, AND NEUTRAL-MESON SYSTEMS%
\footnote{
Presented at Orbis Scientiae, Miami, Florida, January 1997}
\\}
\end{flushleft}

\vglue 0.8cm
\begin{flushleft}
{\hskip 1 truein
V. Alan Kosteleck\'y\\}
\bigskip
{\hskip 1 truein
Physics Department\\}
{\hskip 1 truein
Indiana University\\}
{\hskip 1 truein
Bloomington, IN 47405\\}
{\hskip 1 truein
U.S.A.\\}
\end{flushleft}

\vglue 0.8cm

\noindent
{\bf INTRODUCTION}
\vglue 0.4 cm 

Symmetry under the discrete transformation CPT
is a general theoretical condition holding for
local relativistic field theories of point particles
\cite{cpt1}-\cite{cpt7}.
This symmetry has been investigated experimentally
under different circumstances and to a high degree of precision
\cite{pdg}.
The broad theoretical validity of CPT symmetry for particles
and the availability of high-precision tests
makes CPT violation an interesting candidate experimental signal 
for fundamental theories such as string theory
\cite{kp1,kp2,kp3}.

In this talk,
I consider the possibility that CPT symmetry
might be violated in nature by effects
arising in a theory beyond the standard model.
One example is string theory,
which presently provides the most promising framework 
for a consistent quantum theory of gravity 
incorporating also the known interactions and particles.
Strings are extended objects,
so the usual assumptions underlying proofs of CPT symmetry
do not hold.
Indeed,
spontaneous CPT violation can occur in string theory
\cite{kp1,kp2},
via a mechanism outlined in the next section.

If physics beyond the standard model 
includes spontaneous CPT violation,
it can be described at low energy by additional terms
in an effective theory.
It is possible to establish the general form of such terms 
that are compatible with known gauge symmetries
\cite{kp3,cptsm}.
This analysis in turn
suggests possible consequences of CPT violation
such as baryogenesis
\cite{bckp}
and,
in particular,
quantitative experimental tests of CPT.
Among the most promising tests are those involving 
neutral-meson oscillations,
where specific signatures appear
in experiments involving either correlated or uncorrelated mesons.
These are oulined briefly for the various neutral-meson systems
in subsequent sections.
Further details about these effects and experiments
can be found in the original literature
on the $K$ system
\cite{kp1,kp2,kp3}, 
the two $B$ systems
\cite{kp3,ck1,kv},
the $D$ system
\cite{kp3,ck2}. 

The possible spontaneous CPT violations discussed here,
which are tied to minuscule spontaneous violations 
of Lorentz invariance
\cite{ks},
lie entirely within
the framework of conventional quantum mechanics.
It has also been suggested 
\cite{qg1}-\cite{qg3}
that violations of conventional quantum mechanics 
possibly arising in the context of quantum gravity
might lead to CPT breaking.
The experimental signatures of the two types of CPT
violation in the kaon system are entirely distinct
\cite{qg4}.

\vglue 0.6 cm 
\noindent
{\bf SPONTANEOUS CPT VIOLATION}
\vglue 0.4 cm 

If a fundamental theory underlying nature 
involves more than four spacetime dimensions
and is dynamically Poincar\'e invariant,
some type of spontaneous breaking 
of the higher-dimensional Poincar\'e group 
presumably occurs to generate 
a four-dimensional effective theory.
String theory is most naturally formulated in higher dimensions
and indeed has a mechanism that can trigger spontaneous
Lorentz violation
\cite{ks}.
In string field theory,
this mechanism involves certain interactions that
do not appear in conventional four-dimensional
renormalizable gauge theories.
The string gauge invariance admits these interactions
as a consequence of string nonlocality
or,
equivalently,
as a consequence of the appearance of
an infinite number of particle fields.
If scalar fields in the string theory acquire 
vacuum expectation values,
these interactions can cause destabilizing effects
on the static potentials for Lorentz tensor fields.
A stable vacuum can then become one in which 
Lorentz tensor fields have nonzero expectation values,
thereby spontaneously breaking Lorentz invariance.
If these tensors include ones with an odd number of
spacetime indices,
the spontaneous Lorentz breaking
also involves CPT breaking.

The string field theory of the open bosonic string
provides a useful explicit testing ground for these ideas.
A level-truncation scheme can be used 
to explore the space of extrema for the action
in a systematic way.
The idea is to construct the action and the equations of motion
analytically,
using all particle fields up to a given level number.
The solutions to the equations of motion
that break Lorentz and CPT invariance
can be found
and compared with similar solutions 
for truncations at different level numbers.
Solutions of interest
are those that,
as the level number is increased,
both persist and are corrected by smaller and smaller amounts.
For some situations,
symbolic-manipulation techniques have
enabled us to treat over
20,000 nonvanishing terms in the action.
The Lorentz and CPT properties expected 
from the theoretical mechanism agree 
with those of the solutions found
via the level-truncation approach.

\vglue 0.6 cm 
\noindent
{\bf CPT-VIOLATING EXTENSION TO THE STANDARD MODEL}
\vglue 0.4 cm 

An interesting issue is whether the mechanism 
outlined above could produce breaking of CPT
(and Lorentz invariance) in our four spacetime dimensions.
It would seem natural for this to occur,
since there is no apparent reason why four dimensions
should be preferentially selected in the higher-dimensional theory.
However,
no CPT violation has been experimentally detected,
so any such breaking must be highly suppressed
in the standard model.
In a realistic string theory and
treating the standard model as an effective low-energy model,
the natual dimensionless suppression factor that appears
would be the ratio $r$ of the low-energy scale to the Planck scale,
$r \sim 10^{-17}$.
This suppression factor would produce only a few 
potentially observable CPT-violating effects, 
among which are ones in principle detectable 
in the kaon and other neutral-meson systems 
\cite{kp1,kp3}.

A generic CPT-violating contribution
to the effective four-dimensional low-energy theory
(the standard model)
that could emerge from a compactified string theory
could have the form 
\cite{kp2,kp3}:
\beq
\cl \sim \fr {\la} {M^k} 
\vev{T}\cdot\overline{\ps}\Ga(i\prt )^k\ch
+ {\textstyle h.c.}
\quad .
\label{a}
\eeq
Here, 
$\vev{T}$ is the expectation value of a Lorentz tensor $T$.
The four-dimensional fermions $\ps$ and $\ch$ 
are contracted in spinor space
through a gamma-matrix structure $\Ga$,
with couplings to $T$ possibly involving derivatives $i\prt$.
The factors of the (Planck or compactification) mass $M$ 
must be present on dimensional grounds,
and $\la$ is taken to be a dimensionless coupling constant.

A particularly interesting CPT-violating extension 
of the standard model can be obtained 
from terms of the form \rf{a} 
by identifying the fermions $\ps$ and $\ch$ with ones
appearing in the standard model
and requiring that the usual SU(3) $\times$ SU(2) $\times$ U(1)
gauge invariance is maintained.
The possible terms compatible with
naive power-counting renormalizability 
have been explicitly given in ref.\ \cite{cptsm},
along with a framework for treating theoretically
the accompanying CPT and Lorentz breaking.

The next sections summarize some of the observable
consequences of such terms in neutral-meson systems.
Other effects are also possible.
For example,
under suitable circumstances terms of the form \rf{a}
could produce baryogenesis in thermal equilibrium
\cite{bckp}.
This mechanism for 
generating the observed baryon asymmetry 
is distinct from more conventional ones
that require nonequilibrium processes
and C- and CP-breaking interactions
\cite{ads}.

\vglue 0.6 cm 
\noindent
{\bf NEUTRAL-MESON OSCILLATIONS}
\vglue 0.4 cm 

To investigate possible CPT-violating signals 
in neutral-meson systems,
$\ps$ and $\ch$ can be taken as the quarks comprising the 
neutral meson,
denoted generically by $P$
($P \equiv K$, $D$, $B_d$, or $B_s$).
Terms of the form \rf{a} then produce contributions 
to the $2\times 2$ effective hamiltonian $\La$ governing the 
time evolution of the meson system.
Within the context of conventional quantum mechanics,
there are two kinds of (indirect) CP violation 
that can appear in $\La$:
T-violating contributions that preserve CPT,
and CPT-violating contributions that preserve T.
The corresponding complex parameters are denoted
$\ep_P$ and $\de_P$, respectively.
A plausible theoretical framework for understanding
the appearance of a nonzero value of 
the T-violating parameter $\ep_P$ exists 
in the context of the standard model,
using the CKM matrix.

The CPT-breaking extension of the standard model 
mentioned in the previous section
provides a basis for understanding the origin
of a possible nonzero value of 
the CPT-violating quantity $\de_P$
in terms of spontaneous CPT and Lorentz breaking
as might occur in the string scenario,
for example.
An analysis shows that $\de_P$ can be expressed 
within this framework as 
\cite{kp2,kp3}
\beq
\de_P = i 
\fr{h_{q_1} - h_{q_2}}
{\sqrt{\De m^2 + \De \ga ^2/4}}
e^{i\hat\ph}
\quad .
\label{b}
\eeq
In this equation,
$\De m$ and $\De\ga$
are mass and rate differences
and $\hat\ph = \tan^{-1}(2\De m/\De \ga)$.
These are experimental observables.
The quantities $h_{q_j}=r_{q_j}\la_{q_j}\vev{T}$
originate from the terms \rf{a}
and from the effects of the quark-gluon sea,
parametrized by $r_{q_j}$.

Since the underlying fundamental theory is
assumed to be hermitian
and since the CPT and Lorentz breaking are spontaneous,
the quantities $h_{q_j}$ are real.
This implies the relationship
\beq
\Im \de_P = \pm \cot\hat\ph ~\Re\de_P
\quad , 
\label{c}
\eeq
connecting the real and imaginary parts of $\de_P$
through an experimental observable.
The small size of the suppression ratio $r$
precludes experimental detection
of any direct CPT violation in the decay amplitudes
of the $P$ meson,
so if CPT violation is indeed detected using neutral mesons 
then the result \rf{c} would be the primary signature.

\vglue 0.6 cm
\noindent
{\bf EXPERIMENTAL TESTS}
\vglue 0.4 cm

Experimental tests of CPT violation can be envisaged
in any of the $K$, $D$, $B_d$, and $B_s$
neutral-meson systems
\cite{kp3}.
Within the string-based framework 
described in the previous section,
the CPT-violating parameters $\de_P$ given by \rf{b}
depend on dimensionless coupling constants $\la_{q_j}$
that are presumably of different magnitude
for different quark flavors $q_j$.
This means the $\de_P$ should differ 
for distinct $P$ mesons.
An analogous situation occurs for 
the standard-model Yukawa couplings,
which range over some six orders of magnitude.

One implication of this degree of freedom 
is that CPT symmetry should be
tested experimentally in more than one neutral-meson system.
Another is that some startling possibilities 
might occur in the behavior of heavy neutral mesons.
In the $B_d$ system,
for instance, 
there are currently no bounds on CPT violation
and the bounds on T violation are relatively weak.
It is therefore conceivable that
CPT violation could exceed the expected conventional T violation,
which would produce unexpected signals
in the proposed $B$ factories.

Experiments investigating indirect CP violation
use either uncorrelated neutral mesons $P$
or correlated $P$-$\overline P$ pairs arising
from quarkonium decays.
Typical experimental signatures for CPT and T violation
involve asymmetries of decay probabilities 
into different final states.
Appropriate asymmetries 
with and without time dependence 
and for both correlated and uncorrelated cases
are presently available for all neutral-meson systems. 
These have been used both
for relatively simple theoretical estimates
and as input for detailed Monte-Carlo simulations
of realistic experimental data,
including background effects and acceptances.

In the remainder of this section,
I provide a few remarks about 
the current status of CPT violation 
in the various neutral-meson systems.
The reader is referred to the original literature
\cite{kp1}-\cite{ck2}
for a more complete treatment.

The $K$ system presently offers the only 
neutral-meson limit on CPT violation.
The published bounds 
\cite{pdg,expt1,expt2}
correspond to limits on $|\de_K|$ 
of order $10^{-3}$.
Data from various experiments
recently completed (e.g., CPLEAR at CERN) 
or now underway (e.g., KTeV at Fermilab) 
are likely to lead to an improved bound 
within the near future.

In the $D$ system,
no mixing has yet been observed experimentally
and strong dispersive effects
make theoretical calculations uncertain.
Estimating the CPT reach of future experiments
is therefore relatively difficult.
Nonetheless,
under theoretically favorable circumstances
there are some interesting possibilities for 
placing bounds on $\de_D$ with available techniques
and perhaps even from existing data.

The $B_d$ system is of especial interest for CPT tests
because it involves the heaviest quark  
and so might generate the largest CPT violation.
Currently,
no limit on $\de_{B_d}$ has been published. 
However,
enough data have been obtained to place a bound
on $\de_{B_d}$.
A conservative Monte-Carlo simulation 
with realistic experimental data \cite{kv}
suggests a limit of order 10\% on $\de_{B_d}$
could be extracted by analyzing existing data
from CERN and Cornell.
In any event,
the planned $B$ factories are expected 
to improve this significantly. 

\vglue 0.6 cm
\noindent
{\bf ACKNOWLEDGMENTS}
\vglue 0.4 cm

My thanks to Orfeu Bertolami, Don Colladay, 
Rob Potting, Stuart Samuel, 
and Rick Van Kooten for pleasant collaborations 
leading to results presented in this talk.
This work was supported in part
by the United States Department of Energy 
under grant number DE-FG02-91ER40661.

\vglue 0.6 cm
\noindent
{\bf REFERENCES}
\vglue 0.4 cm

\end{document}